\documentclass[prd,aps,showpacs,nofootinbib,tightenlines]{revtex4}
\usepackage{amsmath}
\usepackage{amssymb}
\usepackage{epsfig}
\usepackage{graphicx}
\usepackage{bm}
\usepackage{times}
\usepackage{braket}
\usepackage{color}
\usepackage{slashed}
\usepackage{hyperref}
\definecolor{Red}{rgb}{1.,0.,0.}

\definecolor{Blue}{rgb}{0.,0.,1.}

\definecolor{nicered}{rgb}{0.7,0.1,0.1}
\definecolor{nicegreen}{rgb}{0.1,0.5,0.1}
\hypersetup{colorlinks,citecolor=nicegreen,linkcolor=nicered}

\newcommand{\beq}{\begin{eqnarray}}
\newcommand{\eeq}{\end{eqnarray}}
\newcommand{\non}{\nonumber\\ }

\newcommand{\calb}{ {\cal B} }

\def \cpc{ Chin. Phys. C  }

\def \epjc{ Eur. Phys. J. C }
\def \ijmpa{ Int. J. Mod. Phys. A }
\def \jpg{  J. Phys. G }
\def \npb{  Nucl. Phys. B }
\def \plb{  Phys. Lett. B }
\def \ppnp{ Prog.Part. $\&$ Nucl. Phys. }
\def \prd{  Phys. Rev. D }
\def \prl{  Phys. Rev. Lett.  }

\def \jhep{ JHEP }

\begin{document}
%%=======================================
\title{Quasi-two-body decays $B \to \eta_c {(1S ,2S)}\;[\rho(770),\rho(1450),\rho(1700) \to ]\; \pi\pi$
in the perturbative QCD approach}
\author{Ya Li$^1$}                \email{liyakelly@163.com}
\author{Ai-Jun Ma$^1$}            \email{theoma@163.com}
\author{Zhou Rui$^2$}              \email{jindui1127@126.com}
\author{Zhen-Jun Xiao$^{1,3}$}    \email{xiaozhenjun@njnu.edu.cn}

%-----------
\affiliation{$^1$ Department of Physics and Institute of Theoretical Physics,
                          Nanjing Normal University, Nanjing, Jiangsu 210023, P.R. China}
\affiliation{$^2$ College of Sciences, North China University of Science and Technology,
                          Tangshan 063009,  China}
\affiliation{$^3$ Jiangsu Key Laboratory for Numerical Simulation of Large Scale Complex
Systems, Nanjing Normal University, Nanjing, Jiangsu 210023, P.R. China}

\date{\today}

%XXXXXXXXXXXXXXXXXXXXXXXXXXXXXXXXXX%
\begin{abstract}
In this paper, we calculated the branching ratios of the quasi-two-body decays
$B \to \eta_c (1S ,2S)$ $[\rho(770), \rho(1450),\rho(1700)\to ] \pi\pi$
by employing the perturbative QCD (PQCD) approach.
The contributions from the $P$-wave resonances $\rho(770)$, $\rho(1450)$ and $\rho(1700)$
were taken into account.
The two-pion  distribution amplitude $\Phi_{\pi\pi}^{\rm P}$ is parameterized by the vector
current time-like form factor $F_{\pi}$ to study the considered decay modes.
We found that (a) the PQCD predictions for the branching ratios of the considered quasi-two-body decays
are in the order of $10^{-7} \sim 10^{-6}$, while the two-body decay rates ${\cal B}(B \to \eta_c{(1S,2S)}
(\rho(1450),\rho(1700)))$ are extracted from those for the corresponding quasi-two-body decays;
(b) the whole pattern of the pion form factor-squared $|F_\pi|^2$ measured by
the {\it BABAR} Collaboration could be understood based on our theoretical results;
(c) the general expectation based on the similarity between $B \to \eta_c \pi\pi$ and
$B \to J/\psi \pi\pi$ decays are confirmed: $R_2(\eta_c)\approx 0.45$ is consistent with the measured $R_2(J/\psi)\approx 0.56\pm 0.09$
within errors;
and (d) new ratios $R_3(\eta_c(1S))$ and $R_4(\eta_c(2S))$ among the branching ratios
of the considered decay modes are defined and could be tested by future experiments.
\end{abstract}


\maketitle

%XXXXXXXXXXXXXXXXXXXXXXXXXXXXXXXXXX% @ Begin

\section{Introduction}

In recent years, due to the great progress in the theoretical studies and experimental measurements, the three-body
hadronic B meson decays become much more attractive than ever before, and begin to play an important role in
testing the standard model (SM) and in searching for the signal of the possible new physics beyond the SM.

In the experiment side, the measurements for the branching ratios and $CP$ violating asymmetries for
$B \to K \pi \pi $ and other decay modes have been reported
by the {\it BABAR} ~\cite{BABAR:01,BABAR:02,BABAR:03,BABAR:04,BABAR:05},  Belle ~\cite{Belle:01,Belle:02,Belle:03,Belle:04}
and LHCb Collaboration~\cite{lhcb0,prl111-101801,jhep10-143,prd90-112004,prl112-011801,prd95-012006,lhcb1702}.
These three-body decays are known experimentally to be dominated by the low energy resonances
on $\pi\pi$, $KK$ and $K\pi$ channels on the Dalitz plots \cite{dalitz-plot1,dalitz-plot2},
analysed by employing the isobar model~\cite{123-333,prd11-3165} in terms of the usual Breit-Wigner
model~\cite{BW-model} or D.V.~Bugg model \cite{Bugg-model} plus a background.
Obviously, such decay modes do receive the resonant and nonresonant contributions, as well as the
possible final-state interactions (FSIs)~\cite{prd89-094013,1512-09284,89-053015},
but the relative strength of these contributions from different sources are varying significantly from channel to channel.

In the theory side, the three-body hadronic decays of the heavy $B$ meson are clearly much more complicated
to be described theoretically than those two-body decays.
We firstly can not separate the nonresonant contributions from the resonant ones clearly,
and  secondly do not know how to calculate or estimate the nonresonant and FSI contributions reliably \cite{ST:15}.
As a first step, however, we can restrict ourselves to specific kinematical configurations, in which two
energetic final state mesons almost collimating to each other, the three-body interactions for such topologies
are expected to be suppressed strongly.
Then it seems reasonable to assume the validity of factorization for these quasi-two-body $B$ decays.
In the ``quasi-two-body" mechanism, the two-body scattering and all possible interactions
between the two involved  particles are included but the interactions between the third
particle and the pair of mesons are neglected.

During the past two decades, several different theoretical frameworks have been developed for the
study of the three-body hadronic
$B$ meson decays: the one based on the QCD-improved factorization (QCDF)
~\cite{plb622-207,prd74-114009,B.E:2009th,ST:15,CY01,CY02,CY16,prd87-076007}, the method with the symmetry principles
~\cite{prd72-094031,plb727-136,prd72-075013,prd84-056002,plb726-337,plb728-579,IJMPA29-1450011,prd91-014029}
and the framework relaying on the perturbative QCD (PQCD) approach
~\cite{Chen:2002th,Chen:2004th,Wang-2014a,Wang-2015a,Wang-2016,epjc76-675,ma16,ly16,zhou17,ma17,ly17}.
In PQCD factorization approach, for example, we study the three-body hadronic decays of B meson
by introducing the two-hadron distribution amplitude (DA) $\Phi_{h_1 h_2}$~\cite{MP,MT01,MT02,MT03,MN,Grozin01,Grozin02}
to describe the system of the two collimating energetic final state mesons.
Our estimation proceeds via  the idea of the quasi-two-body decays involving resonant and nonresonant contributions,
which can be absorbed in the time-like form factors to parameterize these two-hadron DAs.

As discussed in Ref.~\cite{Chen:2002th}, we here assume that the hard $b$-quark decay kernels containing two virtual gluons
at leading order is not important due to the power-suppression.
The contributions from the dynamical region, where there is at least one pair of the final state light mesons
having an invariant mass below $O(\bar\Lambda m_B)$~\cite{Chen:2002th}, $\bar\Lambda=m_B-m_b$ being the $B$ meson
and $b$ quark mass difference, is dominant.
It's reasonable that the dynamics associated with the pair of mesons can be factorized into a two-meson distribution
amplitude $\Phi_{h_1h_2}$.
In the PQCD approach, one can write down the decay amplitude for a $B\to h_1h_2h_3$ decay
symbolically in the following form ~\cite{Chen:2002th}
\begin{eqnarray}
\mathcal{A}=\Phi_B\otimes H\otimes \Phi_{h_1h_2}\otimes\Phi_{h_3},
\end{eqnarray}
where the hard kernel $H$ describes the dynamics of the strong and electroweak
interactions in three-body hadronic decays in a similar way as the one for the two-body $B\to h_1 h_2$ decays,
the function $\Phi_B$ and $\Phi_{h_3}$ are the wave functions for the B meson and the final-state $h_3$ meson.

Up to now, the decays of $B$ mesons to the charmonium state plus a pion pair,
such as the decay modes $B^0\to J/\psi \pi^+\pi^-$~\cite{prl90-091801,prd87-052001,prd90-012003,plb742-38},
$B_s^0\to J/\psi \pi^+\pi^-$~\cite{prd86-052006,prd89-092006}, $B_{(s)}^0\to \psi(2S) \pi^+\pi^-$\cite{npb-871403} and
$B_s^0\to \eta_c \pi^+\pi^-$~\cite{lhcb1702}, have been measured by {\it BABAR} and LHCb Collaboration.
For $\bar{B}^0 \to J/\psi \pi^+\pi^-$ decay ~\cite{prd90-012003}, six interfering $\pi^+\pi^-$ states,
$\rho(770), f_0(500), f_2(1270), \rho^{\prime}(1450), \omega(782)$ and $\rho^{\prime\prime}(1700)$,
are required to give a good description of invariant mass spectra and decay angular distributions.
Along with the rapid progress of the LHCb experiment, more information of the $B$ meson three-body
decays involving various charmonium states ($\eta_c{(1S, 2S)}$ etc.) will become available.
To improve the description of the invariant mass spectra, more resonant structures should be  taken into account.
Very recently, based on the PQCD factorization approach, we studied the $S$-wave resonance contributions to the
decays $B^0_{(s)}\to \eta_c{(1S,2S)}\pi^+\pi^-$ \cite{epjc76-675,ma17} and
$B^0_{s}\to \psi(2s) \pi^+\pi^-$~\cite{zhou17}, as well as the $P$-wave contributions
(i.e. $\rho(770),\rho(1450)$ and $\rho(1700)$ \footnote{ For the sake of simplicity, we
generally use the abbreviation $\rho=\rho(770)$, $\rho^\prime=\rho(1450)$, $\rho^{\prime \prime}
=\rho(1700)$ in the following sections.  } ) to the decays
$B \to P \rho \to P \pi\pi$~\cite{ly16,ly17}.

\begin{figure}[tbp]
%\vspace{-1cm}
\centerline{\epsfxsize=14cm \epsffile{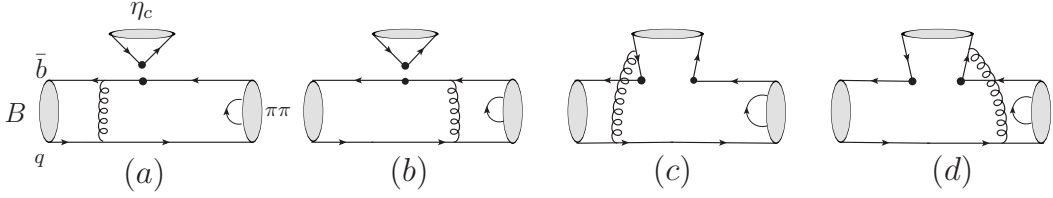}}
%\vspace{0.3cm}
%\vspace{-6cm}
\caption{Typical Feynman diagrams for the quasi-two-body decays $B \to \eta_c{(1S,2S)}(\rho \to) \pi \pi$,
with $q=(u,d)$, and the symbol $\bullet$ denotes the weak vertex.}
\label{fig:fig1}
\end{figure}

In this paper, we will extend our previous analysis to the cases for the $P$-wave resonance
($\rho,\rho^{\prime}$ and $\rho^{\prime\prime}$ ) contributions to
the three-body decays $B \to \eta_c{(1S,2S)} \pi\pi$.
For the quasi-two-body decays $B\to \eta_c{(1S,2S)}
(\rho, \rho^\prime,\rho^{\prime\prime} ) \to \eta_c{(1S,2S)} \pi\pi$, the relevant Feynman diagrams are illustrated
in Fig.~\ref{fig:fig1}. The vector current time-like form factor $F_\pi$~\cite{prl21-244} will be adopted to
describe the strong interactions between the $P$-wave resonant state $(\rho,\rho',\rho'')$ and the final-state
pion pair in our work.
In Sec.~II, we give a brief introduction for the theoretical framework.
The numerical values, some discussions and the conclusions will be given in last two sections.
The explicit PQCD factorization formulas for all the decay amplitudes are collected in the Appendix.

%%%%%%%%%%%%%%%%%%%%%%%%%%%%%%
%%%%%%%%%%%%%%%%%%%%%%%%%%%%%%

\section{Framework}\label{sec:2}  %%%%  Sec-II

For the quasi-two-body $B \to \eta_c{(1S,2S)} (\rho,\rho^\prime,\rho^{\prime\prime}) \to  \eta_c{(1S,2S)}\pi\pi$ decays,
the $B$ meson momentum $p_{B}$, the total momentum of the pion pair $p=p_1+p_2$  and the final-state
$\eta_c$ momentum $p_3$, can be expressed in the light-cone coordinates as the following form:
\begin{eqnarray}\label{mom-pBpp3}
p_{B}=\frac{m_{B}}{\sqrt 2}(1,1,0_{\rm T}),~\quad p=\frac{m_{B}}{\sqrt2}(1-r^2,\eta,0_{\rm T}),~\quad
p_3=\frac{m_{B}}{\sqrt 2}(r^2,1-\eta,0_{\rm T}),
\end{eqnarray}
where $m_{B}$ is the mass of $B$ meson, $\eta=\frac{\omega^2}{(1-r^2)m^2_{B}}$ with $r=m_{\eta_c}/m_{B}$
and the invariant mass squared $\omega^2=p^2$.
In the same way, we also define the momentum $k_B$ of the spectator quark in the $B$ meson,
the momentum $k=z p^+$ and $k_3=x_3 p_3$ for the quark in the resonant state $(\rho,\rho^\prime,\rho^{\prime\prime})$ and
in the final state $\eta_c$ in the following form:
\beq
k_{B}=\left(0,x_B \frac{m_{B}}{\sqrt2} ,k_{B \rm T}\right),\quad
k= \left( z (1-r^2)\frac{m_{B}}{\sqrt2},0,k_{\rm T}\right),\quad
k_3=\left(r^2x_3\frac{m_B}{\sqrt{2}},(1-\eta)x_3 \frac{m_B}{\sqrt{2}},k_{3{\rm T}}\right),
\eeq
where the parameter $x_B, z, x_3$ denotes the momentum fraction of the quark in each meson
and runs from zero to unity.
If we define $\zeta=p^+_1/p^+$ as one of the pion pair's momentum fraction,
other kinematic variables of the two pions can be chosen as
\begin{eqnarray}
p^-_1=(1-\zeta)\eta\frac{m_{B}}{\sqrt2}, \quad p^+_2=(1-\zeta)(1-r^2)\frac{m_{B}}{\sqrt2},
\quad p^-_2=\zeta\eta\frac{m_{B}}{\sqrt2}.
\end{eqnarray}

We assume that the $B \to \eta_c (\rho,\rho^\prime,\rho^{\prime\prime}\to ) \pi\pi$ decays
can proceed mainly via quasi-two-body channels, which contain a $P$-wave resonant state
by introducing the two-pion DAs $\Phi_{\pi\pi}^{\rm P}$.
As done in Ref.~\cite{Wang-2016}, we should introduce the time-like form factor $F_{\pi}(s)$,
which involves the strong interactions between the $P$-wave resonance and two pions, as well
as elastic rescattering of pion pair to parameterize the $P$-wave two-pion distribution amplitudes $\Phi_{\pi\pi}^{\rm P}$.
We adopt the same $F_{\pi}(s)$ in this work as the one in Ref.~\cite{Wang-2016}, the approximate
relations $F_{s,t}(s)\approx (f_\rho^T/f_\rho) F_\pi(s)$~\cite{Wang-2016} will also be used in the following section.
By taking the $\rho-\omega$ interference and the excited states into account, the form factor $F_{\pi}(s)$ can be written in the form of
\beq
F_{\pi}(s)= \left [ {\rm GS}_\rho(s,m_{\rho},\Gamma_{\rho})
\frac{1+c_{\omega} \cdot {\rm BW}_{\omega}(s,m_{\omega},\Gamma_{\omega})}{1+c_{\omega}}
+\sum_i c_i\cdot {\rm GS}_i(s,m_i,\Gamma_i)\right] \left[ 1+\sum_i c_i\right]^{-1}
\label{eq:fpp}
\eeq
where $s=m^2(\pi\pi)$ is the two-pion invariant mass squared,
$i=(\rho(1450), \rho(1700), \rho(2254))$,
$\Gamma_i$ is the decay width for the relevant resonance,
$m_{\rho,\omega,i}$ are the masses of the corresponding mesons, respectively.
The function ${\rm GS}_\rho(s,m_{\rho},\Gamma_{\rho})$ has been parameterized in
the Gounaris-Sakurai (GS) model~\cite{prl21-244} based on the Breit-Wigner (BW) model~\cite{BW-model},
\begin{equation}
{\rm GS}_\rho(s, m_\rho, \Gamma_\rho) =
\frac{m_\rho^2 [ 1 + d(m_\rho) \Gamma_\rho/m_\rho ] }{m_\rho^2 - s + f(s, m_\rho, \Gamma_\rho)
- i m_\rho \Gamma (s, m_\rho, \Gamma_\rho)}~.
\end{equation}
The explicit expressions of the resonant state function $GS_\rho, GS_i $ and $BW_\omega$ and the values
of the involved parameters can be found for example in Ref.~\cite{prd86-032013}.

We here adopt the same two-pion distribution amplitude as the one being used in Ref.~\cite{Wang-2016},
\begin{eqnarray}
\Phi_{\pi\pi}^{\rm P}=\frac{1}{\sqrt{2N_c}}\left [ { p \hspace{-2.0truemm}/ }
\Phi_{v\nu=-}^{I=1}(z,\zeta)+\omega\Phi_{s}^{I=1}(z,\zeta)
+\frac{{p\hspace{-1.5truemm}/}_1{p\hspace{-1.5truemm}/}_2
  -{p\hspace{-1.5truemm}/}_2{p\hspace{-1.5truemm}/}_1}{w(2\zeta-1)}\Phi_{t\nu=+}^{I=1}(z,\zeta)
\right ],
\label{eq:phifunc}
\end{eqnarray}
with
\begin{eqnarray}
\Phi_{v\nu=-}^{I=1}&=&\frac{3F_{\pi}(s)}{\sqrt{2N_c}}z(1-z)\left[1
+a^0_{2\rho}\cdot\frac{3}{2}\left[ 5(1-2z)^2-1 \right]\right] P_1(2\zeta-1) \;,\\
\Phi_{s}^{I=1}&=&\frac{3F_s(s)}{2\sqrt{2N_c}}(1-2z)\left[1
+a^s_{2\rho}\cdot \left ( 10z^2-10z+1 \right )\right] P_1(2\zeta-1) \;,\\
\Phi_{t\nu=+}^{I=1}&=&\frac{3F_t(s)}{2\sqrt{2N_c}}(1-2z)^2\left[1
+a^t_{2\rho}\cdot \frac{3}{2}\left[ 5(1-2z)^2-1 \right] \right] P_1(2\zeta-1) \;,
\end{eqnarray}
where the Legendre polynomial $P_1(2\zeta-1)=2\zeta-1$.
In the numerical calculations, we will use the same set of Gegenbauer moments $a^{0,s,t}_{2\rho}$
in the two-pion distribution amplitude $\Phi_{\pi\pi}^{\rm P}$ as those used in Refs.~\cite{ly16,ly17},
\begin{eqnarray}
a^0_{2\rho}=0.30\pm 0.05, \quad a^s_{2\rho}=0.70\pm 0.20, \quad a^t_{2\rho}=-0.40\pm 0.10.
\end{eqnarray}

\section{Numerical results and discussions}\label{sec:3}  %%% SEC-3

The following input parameters (in units of  GeV) will be adopted~\cite{pdg2016} for numerical calculations,
\begin{eqnarray}
\Lambda^{4}_{ \overline{MS} }&=&0.25, \quad m_{B^{\pm,0}}=5.28, \quad
m_{b}=4.8, \quad m_c=1.275 \pm 0.025, \quad m_\rho=0.775, \quad \Gamma_\rho=0.149, \nonumber\\
m_{\pi^\pm}&=&0.140, \quad m_{\pi^0}=0.135,
\quad m_{\eta_c(1S)}=2.9834, \quad m_{\eta_c(2S)}=3.6392, \quad f_B= 0.19\pm 0.02. \label{eq:inputs}
\end{eqnarray}
The values of the Wolfenstein parameters are the same as given in Ref.~\cite{pdg2016}:
$A=0.811\pm0.026, \lambda=0.22506\pm 0.00050$, $\bar{\rho} = 0.124^{+0.019}_{-0.018}$, $\bar{\eta}= 0.356\pm 0.011$.

For the decay $B \to \eta_c (\rho \to \pi \pi)$, the differential decay rate is written as
\begin{eqnarray}
\frac{d{\cal B}}{ds}=\tau_{B}\frac{|\overrightarrow{p_1}||\overrightarrow{p_3}|}{32\pi^3m^3_{B}}|{\cal A}|^2, \label{expr-br}
\end{eqnarray}
with the kinematic variables $|\overrightarrow{p_1}|$ and $|\overrightarrow{p_3}|$
\begin{eqnarray}
|\overrightarrow{p_1}|=\frac{1}{2}\sqrt{s-4m^2_{\pi}}, \quad~~
|\overrightarrow{p_3}|=\frac12  \sqrt{\big[(m^2_{B}-m_{\eta_c}^2)^2-2(m^2_{B}+m_{\eta_c}^2) s
 +s^2 \big]/s},
 \label{br-momentum}
\end{eqnarray}
where $\tau_{B^\pm}=1.638\;{\rm ps}, \tau_{B^0}=1.520\; {\rm ps}$ is the mean lifetime of
$B^\pm$ and $B^0$ meson.

By using the differential decay rate as defined in Eq.~(\ref{expr-br}) and the relevant decay amplitudes
as given in the Appendix, we make the PQCD predictions for the branching rations
$\calb(B \to \eta_c{(1S,2S)} (\rho,\rho^\prime,\rho^{\prime\prime} \to )\pi \pi) $ and find the following
numerical results (in units of $10^{-6}$ )
\begin{eqnarray}
\mathcal {B}(B^+ \to \eta_c(1S)(\rho^+\to)\pi^+ \pi^0 )&=&8.55^{+3.92}_{-2.55}(\omega_B)^{+1.85}_{-1.18}
(a^t_{2\rho})^{+1.58}_{-1.31}(a^s_{2\rho})^{+0.57}_{-0.29}(a^0_{2\rho})\;,
\non
\mathcal {B}(B^+ \to \eta_c(1S)(\rho^{\prime+}\to)\pi^+ \pi^0)&=&
0.93^{+0.23}_{-0.19}(\omega_B)^{+0.07}_{-0.01}(a^t_{2\rho})^{+0.14}_{-0.13}(a^s_{2\rho})\pm{0.02}(a^0_{2\rho})^{+0.26}_{-0.22}(c_{\rho^{\prime}})\;, \non
\mathcal {B}(B^+ \to \eta_c(1S)(\rho^{\prime\prime+}\to)\pi^+ \pi^0)&=&
0.24^{+0.04}_{-0.03}(\omega_B)^{+0.04}_{-0.00}(a^t_{2\rho})^{+0.05}_{-0.04}(a^s_{2\rho})^{+0.01}_{-0.00}
(a^0_{2\rho})^{+0.07}_{-0.06}(c_{\rho^{\prime\prime}}) \;,
\label{eq:br11}\\
\mathcal {B}(B^0 \to \eta_c(1S)(\rho^0\to)\pi^+ \pi^-)&=&
3.95^{+1.85}_{-1.16}(\omega_B)^{+0.86}_{-0.53}(a^t_{2\rho})^{+0.74}_{-0.59}(a^s_{2\rho})^{+0.27}_{-0.13}(a^0_{2\rho})\;, \non
\mathcal {B}(B^0 \to \eta_c(1S)(\rho^{\prime 0}\to)\pi^+ \pi^-)&=&0.43^{+0.10}_{-0.09}(\omega_B)
^{+0.03}_{-0.01}(a^t_{2\rho})\pm{0.06}(a^s_{2\rho})\pm{0.01}(a^0_{2\rho})^{+0.13}_{-0.10}(c_{\rho^{\prime}})\;, \non
\mathcal {B}(B^0 \to \eta_c(1S)(\rho^{\prime\prime0}\to)\pi^+ \pi^-)&=&0.11^{+0.02}_{-0.01}(\omega_B)
^{+0.02}_{-0.00}(a^t_{2\rho})\pm{0.02}(a^s_{2\rho})\pm{0.00}(a^0_{2\rho})\pm{0.03}(c_{\rho^{\prime\prime}})\;.
\label{eq:br12}
\end{eqnarray}
For the decays $B \to \eta_c(1S) (\rho \to )\pi \pi$, the first error of the PQCD predictions comes from the
uncertainty of $\omega_B=(0.40 \pm 0.04)$ {\rm GeV}, the following three errors are due to
$a^t_{2\rho}=-0.40 \pm 0.10$, $a^s_{2\rho}=0.70 \pm 0.20$ and $a^0_{2\rho}=0.30 \pm 0.05$ respectively.
For the decay modes involving $\rho'$ and $\rho''$ resonant states, the fifth error results from the
uncertainty of the form factor $F_{\pi}(s)$ as given in Eq.~({\ref{eq:fpp}}): the one induced by the uncertainties of
the coefficients $c_{\rho^{\prime}}=(0.158\pm0.018)\cdot \exp[i(3.76\pm0.10)]$
and $c_{\rho^{\prime\prime}}= (0.068\pm0.009 )\cdot \exp[i ( 1.39\pm0.20)]$ ~\cite{prd86-032013}.
One can see from the PQCD predictions as given in Eqs.~(\ref{eq:br11},\ref{eq:br12}) that the major error in our
approach comes from the parameter $\omega_B$ in $B$ meson wave function, which  can reach $30-50 \%$.
The error from the coefficient $c_{\rho^{\prime}}(c_{\rho^{\prime\prime}})$ is around $20-30 \%$
for the relevant decay modes.
The possible errors due to the uncertainties of $m_c$ and CKM matrix elements are very small and
can be neglected safely.

For the considered decay modes $B \to \eta_c(1S,2S) \pi\pi$ decay, the dynamical limit on the
value of invariant mass $\omega$ is $2m_\pi \leq \omega \leq ( m_B-m_{\eta_c(1S,2S)} )$.
For $B \to \eta_c(2S) \pi\pi$ decays, since $m(\rho^{\prime\prime}) > \omega_{max}=(m_B-m_{\eta_c(2S)})$,
the resonant $\rho^{\prime\prime}$ can not contribute to this decay.
We therefore have the following PQCD predictions for the branching ratios ( in units of $10^{-6}$ ):
\begin{eqnarray}
\mathcal {B}(B^+ \to \eta_c(2S)(\rho^+\to)\pi^+ \pi^0 )&=&3.82^{+1.45}_{-1.01}(\omega_B)
^{+0.49}_{-0.44}(a^t_{2\rho})^{+0.58}_{-0.55}(a^s_{2\rho})\pm0.14(a^0_{2\rho}) \;, \non
\mathcal {B}(B^+ \to \eta_c(2S)(\rho^{\prime+}\to)\pi^+ \pi^0)&=&0.15\pm{0.03}(\omega_B)
\pm{0.01}(a^t_{2\rho})\pm{0.02}(a^s_{2\rho})\pm 0.00 (a^0_{2\rho})\pm{0.04}(c_{\rho^{\prime}})\;,
\label{eq:br21}\\
\mathcal {B}(B^0 \to \eta_c(2S)(\rho^0\to)\pi^+ \pi^-)&=& 1.77^{+0.68}_{-0.47}(\omega_B)
^{+0.23}_{-0.21}(a^t_{2\rho})^{+0.27}_{-0.25}(a^s_{2\rho})\pm0.06(a^0_{2\rho}) \;, \non
\mathcal {B}(B^0 \to \eta_c(2S)(\rho^{\prime0}\to)\pi^+ \pi^-)&=&0.07\pm 0.01(\omega_B)
\pm 0.01 (a^t_{2\rho})\pm 0.01 (a^s_{2\rho}) \pm 0.00 (a^0_{2\rho})\pm{0.02}(c_{\rho^{\prime}})\;.
\label{eq:br22}
\end{eqnarray}
The errors in above equations have the same meaning as those in Eqs.(\ref{eq:br11},\ref{eq:br12}).

For the phenomenological study of the two-body decays $B \to \eta_c \rho^{\prime}$ and $B \to \eta_c \rho^{\prime\prime}$,
we currently still lack the distribution amplitudes of the states $\rho^{\prime}$ and $\rho^{\prime\prime}$.
But one can extract out the branching fractions for the two-body decays $B \to \eta_c \rho^{\prime} (\rho^{\prime\prime})$
from those PQCD predictions for the quasi-two-body processes $B \to \eta_c \rho^{\prime} (\rho^{\prime\prime})
\to \eta_c \pi\pi$ with the input of $\Gamma_{\rho^\prime \to \pi\pi}/\Gamma_{\rho^\prime}$ and
$\Gamma_{\rho^{\prime\prime} \to \pi\pi}/\Gamma_{\rho^{\prime\prime}}$.
We know that there is a relation of the decay rates between the quasi-two-body and the corresponding
two-body decay modes
\begin{eqnarray}
\mathcal{B}( B \to \eta_c (\rho^{\prime}(\rho^{\prime\prime}) \to ) \pi \pi ) =
\mathcal{B}( B \to \eta_c \rho^{\prime}(\rho^{\prime\prime})) \cdot {\mathcal B}(\rho^{\prime}(\rho^{\prime\prime}) \to\pi\pi).
\label{eq:def1}
\end{eqnarray}
If we take the values ${\mathcal B}(\rho^\prime\to\pi\pi)=10.04^{+5.23}_{-2.61}\%$
and ${\mathcal B}(\rho^{\prime\prime} \to \pi\pi)=8.11^{+2.22}_{-1.47}\%$ as estimated in Ref.~\cite{ly17}
as our input, we can find the PQCD predictions for  $ {\cal B}( B \to \eta_c \rho^{\prime} )$ and
$ {\cal B}( B \to \eta_c \rho^{\prime\prime})$ out of those as given in Eqs.~(\ref{eq:br11}-\ref{eq:br22}):
\begin{eqnarray}
\mathcal {B}(B^+ \to \eta_c(1S)\rho^{\prime+})&=&[9.27^{+3.80}_{-3.18}]\times 10^{-6}\;, \non
\mathcal {B}(B^+ \to \eta_c(1S)\rho^{\prime\prime+})&=&[2.97^{+1.28}_{-0.96}]\times 10^{-6}\;, \non
\mathcal {B}(B^+ \to \eta_c(2S)\rho^{\prime+})&=&[1.47\pm{0.55}]\times 10^{-6}\;,
\label{eq:2bbr1}\\
\mathcal {B}(B^0 \to \eta_c(1S)\rho^{\prime0})&=&[4.32^{+1.77}_{-1.47}]\times 10^{-6}\;, \non
\mathcal {B}(B^0 \to \eta_c(1S)\rho^{\prime\prime0})&=&[1.38^{+0.57}_{-0.46}]\times 10^{-6}\;,\non
\mathcal {B}(B^0 \to \eta_c(2S)\rho^{\prime0})&=&[0.69\pm0.26]\times 10^{-6}\;. \label{eq:2bbr2}
\end{eqnarray}
Here the individual errors from different sources have been added in quadrature.

%%%%%%%%%%%%%%%%%%  Fig-1
\begin{figure}[tbp]
%\vspace{-0.5 cm}
\centerline{\epsfxsize=8.5cm \epsffile{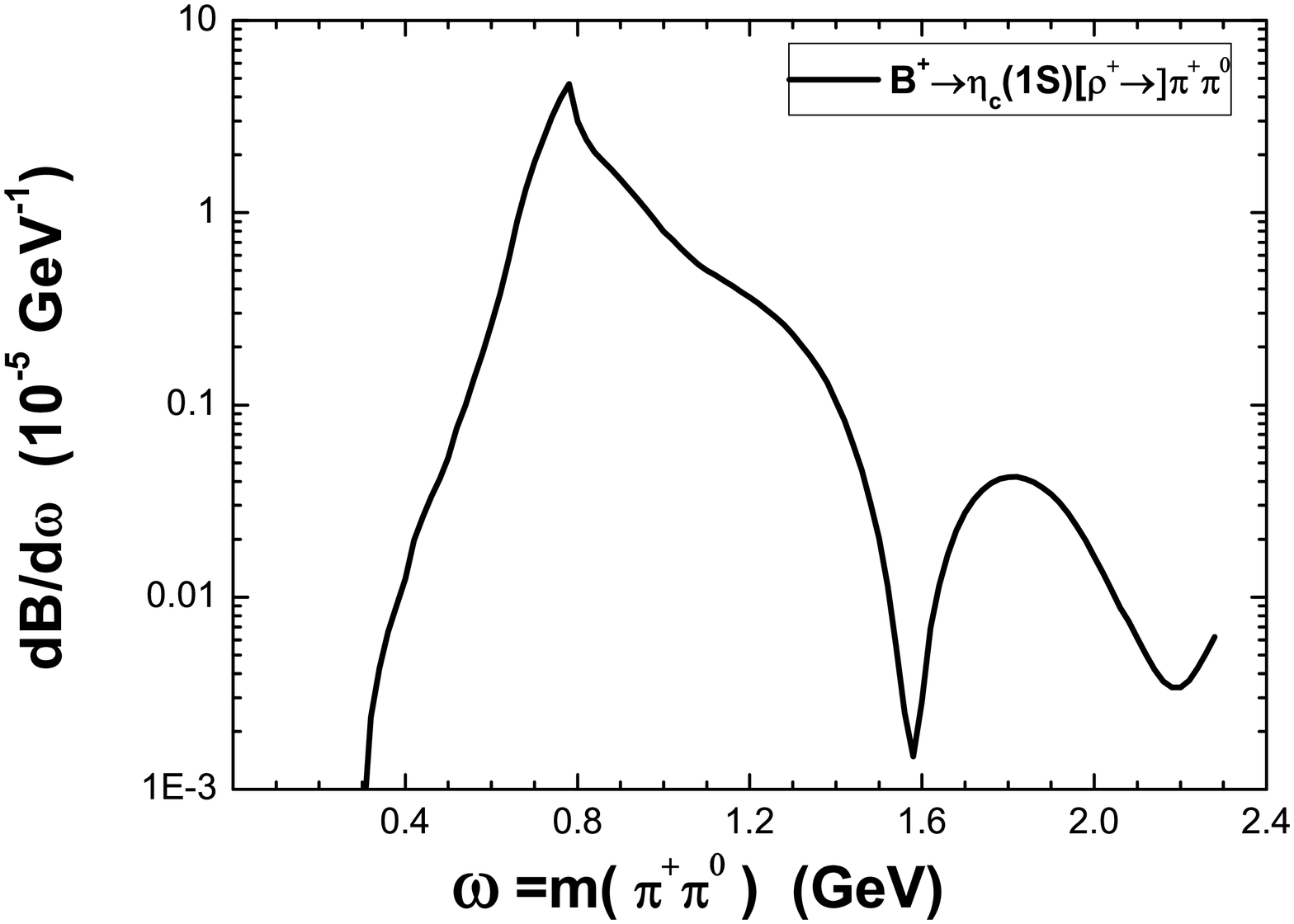}
            \epsfxsize=8.5cm \epsffile{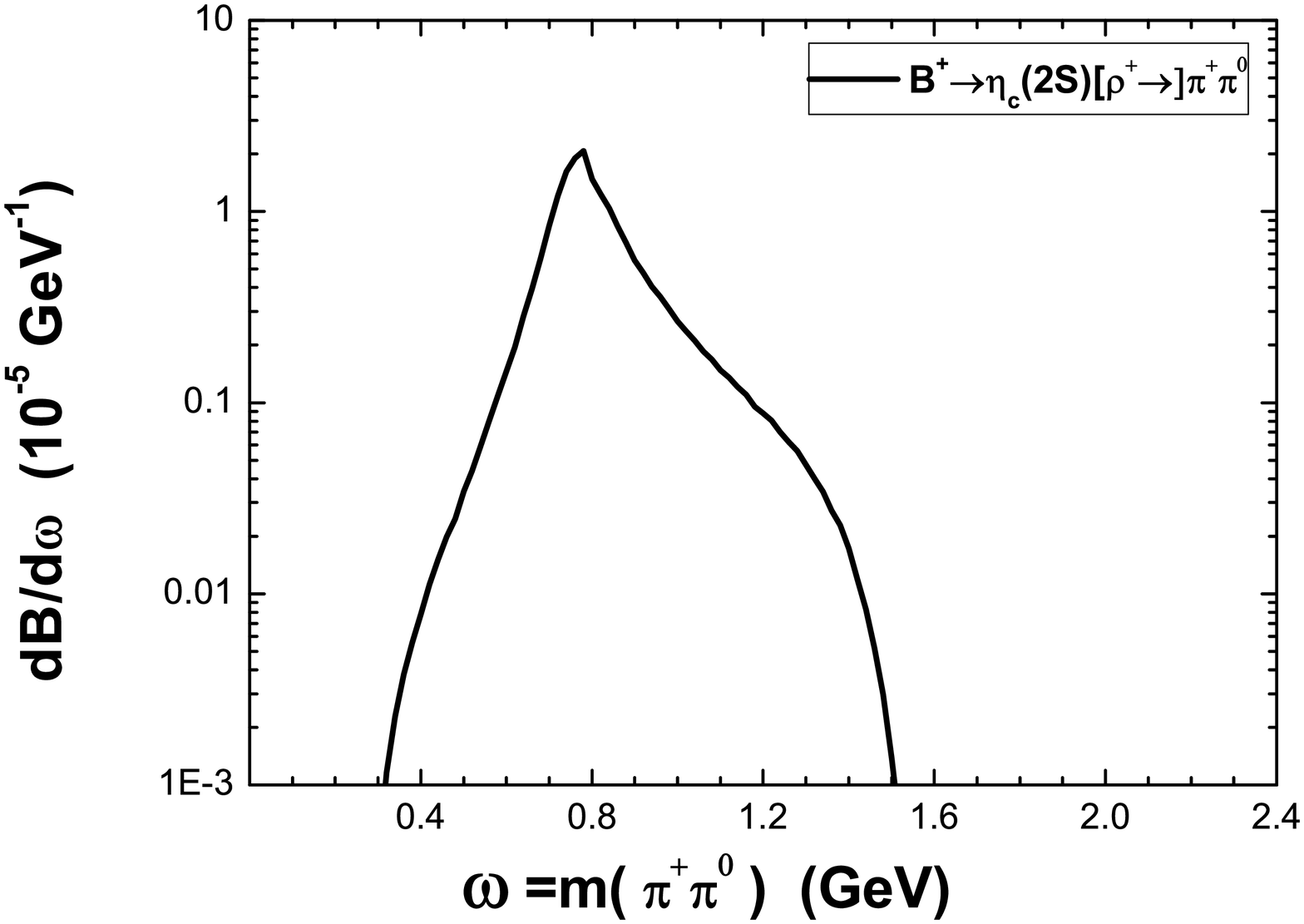}}
\vspace{-0.2cm}
  {\scriptsize\bf (a)\hspace{8.5cm}(b)}
\caption{(a) The PQCD prediction for the differential decay rate of $B^+\to \eta_c(1S)(\rho^+ \to)\pi^+\pi^0$ decay with the
inclusion of all  contributions from $\rho(770)$, $\rho(1450)$ and $ \rho(1700)$.
(b) The differential decay rate of $B^+\to \eta_c(2S)(\rho^+ \to)\pi^+\pi^0$ decay when the possible
contributions from both $\rho(770)$ and $\rho(1450)$ are included.}
\label{fig2}
\end{figure}
%%%%%%%%%%%%%%%%%% Fig-1

\begin{figure}[tbp]
%\vspace{-0.5 cm}
\centerline{\epsfxsize=8.5cm \epsffile{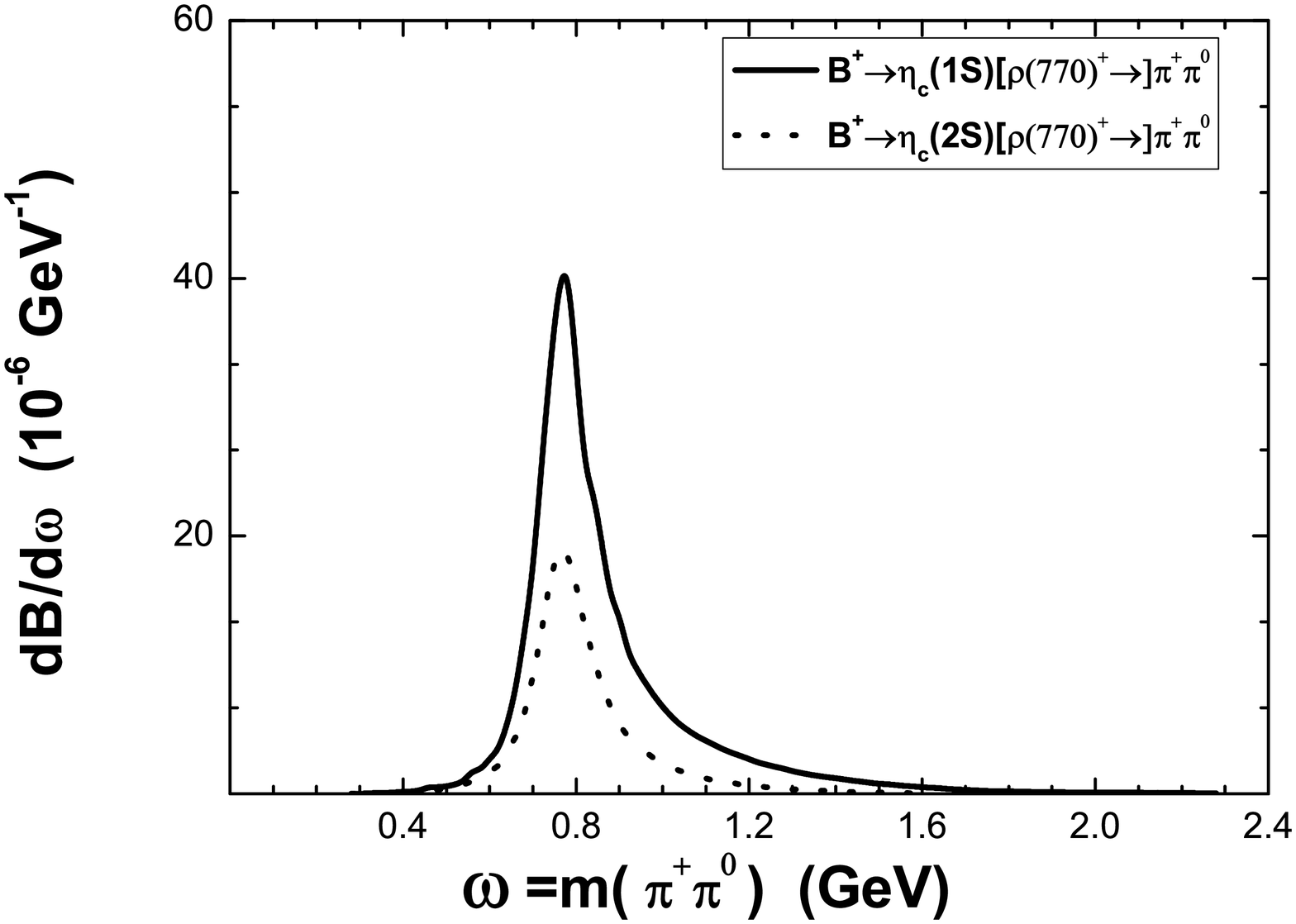}
            \epsfxsize=8.5cm \epsffile{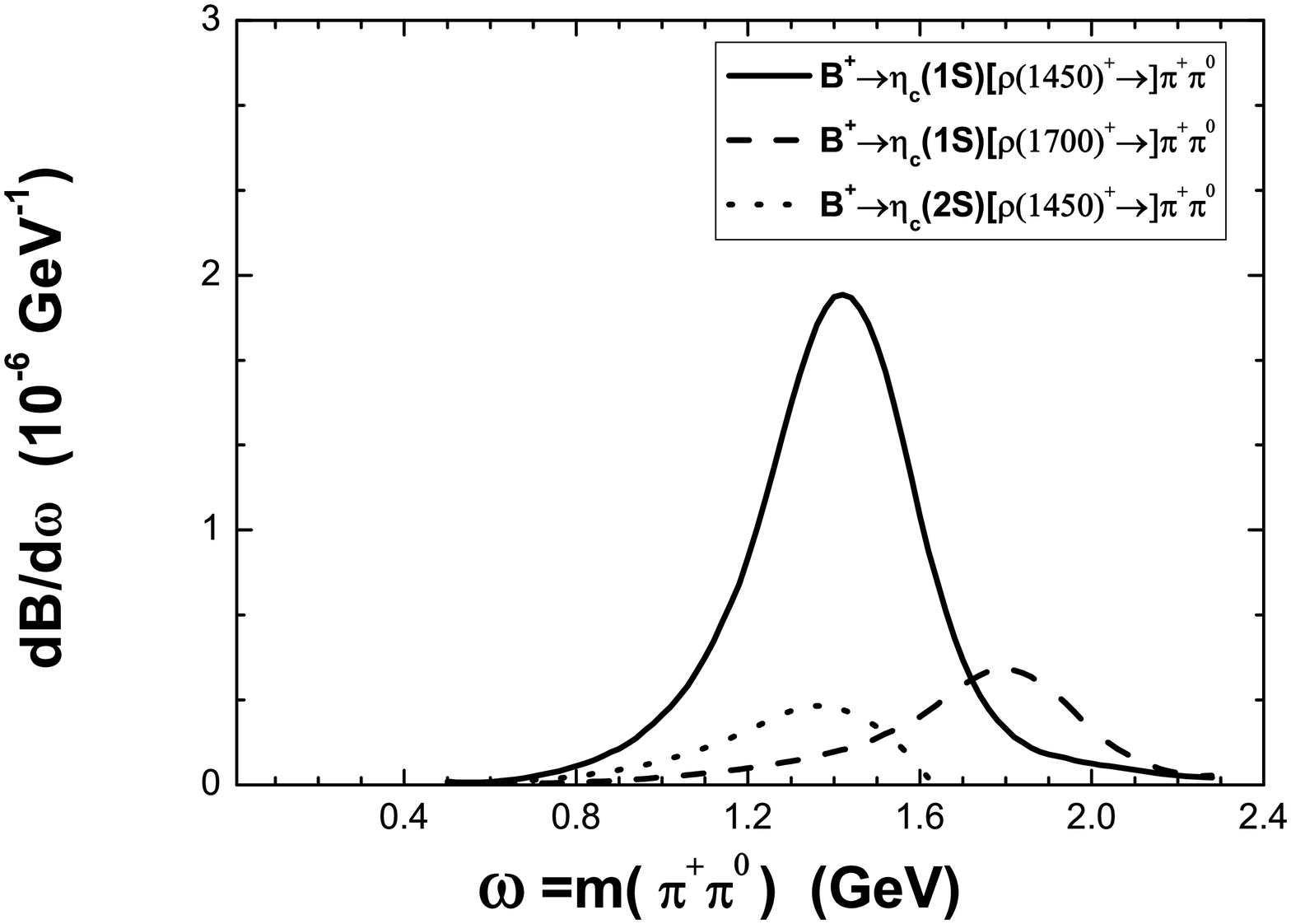}}
\vspace{-0.2cm}
  {\scriptsize\bf (a)\hspace{8.2cm}(b)}
\caption{The PQCD predictions for $d{\cal B}/d\omega$ for
$B^+\to \eta_c(1S)(\rho(770)^+ \to)\pi^+\pi^0$  and other four decay modes.}
\label{fig3}
\end{figure}

In Fig.~\ref{fig2}(a) and \ref{fig2}(b), we show the $\omega$-dependence of the differential decay rate
$d{\cal B}(B^+ \to \eta_c(1S) \pi^+\pi^0)/d\omega$ and $d{\cal B}(B^+ \to \eta_c(2S) \pi^+\pi^0)/d\omega$
after the inclusion of the possible contributions from the resonant states.
For $B^+ \to \eta_c(1S) \pi^+\pi^0$ decay, the dynamical limit is $0.28 {\rm GeV} \leq \omega \leq 2.28 {\rm GeV}$:
all three resonant states $(\rho, \rho^\prime, \rho^{\prime\prime})$ can contribute.
For $B^+ \to \eta_c(2S) \pi^+\pi^0$ decay, however, the limit is $0.28 {\rm GeV} \leq \omega \leq 1.60 {\rm GeV}$:
which means that the heavier $\rho^{\prime\prime}$ can not contribute to this decay mode.

In Fig.~\ref{fig3}(a) and \ref{fig3}(b), we show the $\omega$-dependence of the differential decay rate
of the five considered decay modes.
In Fig.~\ref{fig3}(a), we show the PQCD prediction for $d{\cal B}/d\omega$ for
$B^+\to \eta_c(1S)(\rho(770) \to)\pi^+\pi^0$ decay ( the solid curve) and
$B^+\to \eta_c(2S)(\rho(770) \to)\pi^+\pi^0$ decay (the dotted curve), respectively.
In Fig.~\ref{fig3}(b), similarly, we show the PQCD prediction for $d{\cal B}/d\omega$ for
$B^+\to \eta_c(1S)(\rho(1450) \to)\pi^+\pi^0$ decay ( the solid curve),
$B^+\to \eta_c(1S)(\rho(1700) \to)\pi^+\pi^0$ decay ( the short-dashed curve)
and $B^+\to \eta_c(2S)(\rho(1450)\to)\pi^+\pi^0$ decay (the dotted curve), respectively.

From the curves as illustrated in Fig.~\ref{fig2} and Fig.~\ref{fig3} and the PQCD predictions for the decay rates
as given in Eqs.~(\ref{eq:br11}-\ref{eq:2bbr2}), we have the following observations:
\begin{itemize}
\item[(1)]
According to the full Dalitz-plot analysis to the $B \to J/\psi \pi^+\pi^-$ decay by the
LHCb experiment ~\cite{prd90-012003},  the dominant contributions come from the $P$-wave
resonance $\rho(770)$ and $S$-wave resonance $f_0(500)$.
The relative rate between the two contributions was measured to be in the range
\beq
1.4 \leq R_{J/\psi}\approx \frac{{\cal B}(B^0 \to J/\psi(\rho(770) \to) \pi^+\pi^-)}{{\cal B}(B^0
\to J/\psi(f_0(500) \to) \pi^+\pi^-)} \leq 1.9,
\eeq
here only the fraction of the helicity $\lambda=0$ component of the $P$-wave resonance has been taken into account.
Because of the analogous properties of the $\eta_c$ and $J/\psi$ meson, it is reasonable for us to expect
a similar invariant mass distribution for $B \to \eta_c \pi^+\pi^-$ decay when compared with that of
the $B \to J/\psi \pi^+\pi^-$ decay.

In a previous work ~\cite{epjc76-675}, we calculated the $S$-wave resonance contributions to
$B^0 \to \eta_c(1S) \pi^+\pi^-$ decay, and confirmed that the largest contribution is from the $f_0(500)$.
The PQCD predictions for the branching ratios are
\beq
{\cal B}(B^0\to \eta_c(1S) (f_0(500) \to)\pi^+\pi^- )=  \Big \{ \begin{array}{ll}
1.53 ^{+0.76}_{-0.35}  \times 10^{-6}\;, & {\rm in \ \ BW\ \ model\; [21]\;,} \\
2.31 ^{+0.96}_{-0.48} \times 10^{-6}\;,  & {\rm in \ \ Bugg\ \ model\; [22]\;. } \\
\end{array}
\label{eq:f0500}
\eeq
By using the PQCD prediction as given in Eq.~(\ref{eq:br12}), one can define
the relative ratio of the $P$-wave and $S$-wave contribution as the following form:
\beq
R_1&=&\frac{{\cal B}(B^0 \to \eta_c(1S)(\rho(770) \to) \pi^+\pi^-)}{{\cal B}(B^0
\to \eta_c(1S)(f_0(500) \to) \pi^+\pi^-)}
\approx  \Big \{ \begin{array}{ll}
2.6\;, & {\rm in \ \ BW\ \ model\;  [21] \;,} \\
1.7\;, & {\rm in \ \ Bugg\ \ model\;  [22]\;. } \\ \end{array}
\label{eq:ratioa}
\eeq
The ratio $R_1$ agrees well with  the ratio $R_{J/\psi}$ for the case of
$B \to J/\psi \pi^+\pi^-$ decay and will be tested by the future LHCb and Belle-II experiment.

\item[(2)]
From Fig.~\ref{fig2}(a), one can see one prominent $\rho$ peak, a shoulder around the $\rho(1450)$
and a deep dip near $\omega\approx 1.6$ GeV, followed by an enhancement (the second lower but wider peak )
in the $\rho(1700)$ region.
Because the differential decay rate  $d{\cal B}/d\omega$ depends on the values of $|F_\pi|^2$,
the position of the first peak and deep dip, as well as the pattern of the whole
curve do agree well with the curve in Fig.~45 of the Ref.~\cite{prd86-032013}, where the pion form
factor-squared  $|F_\pi|^2$ measured by ${\it BABAR}$ are illustrated as a function of
$\sqrt{s^\prime}$ (i.e.$m(\pi\pi)$) in the region from $0.3$ to $3$ {\rm GeV}.

The first dip around $\omega \approx 1.6~{\rm GeV}$ is in fact caused by the strong
destructive interference between the resonant state $\rho(1450)$ and $\rho(1700)$.
Taking $B^+ \to \eta_c(1S)\rho(1450) \to \eta_c(1S)\pi^+\pi^0 $ and
$B^+ \to \eta_c(1S) \rho(1700) \to \eta_c(1S)\pi^+\pi^0 $ decay as an example,
we calculated the interference terms between $\rho(1450)$ and $\rho(1700)$ amplitudes and
found the large negative contribution to the total branching ratio.
Numerically, the PQCD predictions for the individual decay rate and the interference term are:
\beq
{\cal B}(B^+ \to \eta_c(1S)(\rho(1450)\to ) \pi^+\pi^0) &\approx & 9.31\times 10^{-7}\;, \non
{\cal B}(B^+ \to \eta_c(1S)(\rho(1700) \to ) \pi^+\pi^0) &\approx & 2.41\times 10^{-7}\;, \non
{\rm interference \ \  term} &\approx &  -6.45\times 10^{-7}\;.
\label{eq:inter}
\eeq
By comparing with other two individual contributions, we find that the interference term is
indeed large and negative, which leads to the first deep dip in the region around $\omega \approx 1.6$ GeV, as illustrated
in Fig.~\ref{fig2}(a).

\item[(3)]
From Fig.~\ref{fig3}(a),  one can see easily that the differential decay rate $d{\cal B}/d\omega$
for $B^+ \to \eta_c(2S)(\rho(770) \to) \pi^+\pi^0)$ decay is always smaller than that
for $B^+ \to \eta_c(1S)(\rho(770) \to) \pi^+\pi^0)$ decay, mainly due to the difference between the
distribution amplitudes of the $\eta_c(1S)$ and $\eta_c(2S)$: the tighter phase space and the smaller decay
constant of the $\eta_c(2S)$ state result in the suppression as shown in Fig.~\ref{fig3}(a).

From the numerical results as given in Eqs.~(\ref{eq:br11}-\ref{eq:br22}),
we obtain the relative ratio $R_2$ between the branching ratios of $B$ meson decays involving
$\eta_c(2S)$ and $\eta_c(1S)$ respectively,
\beq
R_2(\eta_c)=\frac{{\cal B}(B^+ \to \eta_c(2S)(\rho(770) \to) \pi^+\pi^0)}{{\cal B}(B^+ \to \eta_c(1S)(\rho(770) \to)
\pi^+\pi^0)}=\frac{{\cal B}(B^0 \to \eta_c(2S) (\rho(770) \to) \pi^+\pi^-)}{{\cal B}(B^0
\to \eta_c(1S)(\rho(770) \to) \pi^+\pi^-)}\approx 0.45.
\eeq
Owing to the same quark structures between $\eta_c$ and $J/\psi$ mesons, one generally expect that
the $B \to \eta_c \pi\pi$ decays should be similar in nature with the decays $B \to J/\psi \pi\pi$:
i.e. $R_2(\eta_c) \approx R_2(J/\psi)$.
This general expectation, in fact, agrees well with the LHCb measurement \cite{npb-871403}:
\beq
R_2(J/\psi)|_{\rm LHCb} = \frac{{\cal B}(B^0 \to \psi(2S) \pi^+\pi^-)}{{\cal B}(B^0 \to J/\psi \pi^+\pi^-)}
= 0.56 \pm 0.09.
\eeq
Here the main contribution also come from $B^0 \to J/\psi \rho(770) \to J/\psi \pi^+\pi^-$.

\item[(4)]
From Fig.~\ref{fig3}(b),  one can see easily that the differential decay rate $d{\cal B}/d\omega$
for $B^+ \to \eta_c(2S)(\rho(1450) \to) \pi^+\pi^0$ and $B^+ \to \eta_c(1S)(\rho(1700) \to) \pi^+\pi^0$
decay  are much smaller than that for $B^+ \to \eta_c(1S)(\rho(1450) \to) \pi^+\pi^0$ decay.
From the numerical results as given in Eqs.~(\ref{eq:br11}-\ref{eq:br22}), we find the following relative ratios
\beq
R_3(\eta_c(1S))=\frac{{\cal B}(B^+ \to \eta_c(1S) [\rho(1700)\to ]\pi^+\pi^0)}{
{\cal B}(B^+ \to \eta_c(1S) [\rho(1450)\to ]\pi^+\pi^0) }\approx 0.26\;, \\
R_4(\eta_c(2S))=\frac{{\cal B}(B^+ \to \eta_c(2S) [\rho(1450)\to ]\pi^+\pi^0)}{
{\cal B}(B^+ \to \eta_c(1S) [\rho(1450)\to ]\pi^+\pi^0) }\approx 0.16\;.
\eeq
The ratio $R_3(\eta_c(1S))$ is mainly governed by the difference between
the parameters $(c_{\rho^\prime}, c_{\rho^{\prime\prime}})$ and the functions
$GS_{\rho^\prime}$ and $GS_{\rho^{\prime\prime}}$, while the ratio $R_4(\eta_c(2S))$
has a strong dependence on the distribution amplitudes of the $\eta_c(1S)$ and $\eta_c(2S)$.

Based on the similarity between $(\eta_c(1S),\eta_c(2S))$ and $(J/\psi,\psi(2S))$ mesons, furthermore,
it also be reasonable for us to expect similar
$R_3$ and $R_4$ ratios for the cases of $B \to J/\psi \pi \pi$ and $B \to \psi(2S) \pi \pi$ decays.
Fortunately, the ratio $R_3(J/\psi)$ analogous to $R_3(\eta_c(1S))$  has been measured by LHCb
Collaboration recently \cite{prd90-012003}.
If we take only the contributions from the longitudinal component $\rho(1450)_0$ and $\rho(1700)_0$
into account, we can obtain the value of the ratio $R_3(J/\psi)$ from the ``Fit fractions of contributing components"
as listed in Table VI of Ref.~\cite{prd90-012003}:
\beq
R_3(J/\psi)=\frac{{\cal B}(B^0 \to J/\psi [\rho(1700)_0\to ]\pi^+\pi^-)}{
{\cal B}(B^0 \to J/\psi [\rho(1450)_0 \to ]\pi^+\pi^-) } \approx 0.29 \pm 0.16\;, {\rm in\ \ Best-Model},
\eeq
which indeed agrees very well with $R_3(\eta_c(1S))\approx 0.26$. Other predictions
will be tested by the forthcoming LHCb and Belle-II experimental measurements.

\item[(5)]
For $B^+ \to \eta_c(1S) [\rho(770)\to ]\pi^+\pi^0$ decay, the main portion of the branching ratios
lies in the region around the pole mass of $\rho(770)$ meson, as
can be seen clearly in Fig.~\ref{fig3}(a).
The central values of the branching ratio ${\cal B}$ are $4.6 \times 10^{-6}$ and $6.4\times 10^{-6}$
when the integration over $\omega$ is limited in the range of
$\omega=[m_\rho-0.5\Gamma_\rho, m_\rho+0.5\Gamma_\rho]$ or
$\omega=[m_\rho-\Gamma_\rho, m_\rho+\Gamma_\rho]$ respectively, which amount to
$54\%$ and $75\%$ of the total branching ratio ${\cal B}=8.6\times10^{-6}$ as listed in Eq.~(\ref{eq:br11}).

\end{itemize}

\section{CONCLUSION}

In this work, we studied the contributions from the $P$-wave resonance $\rho(770)$,
$\rho(1450)$ and $\rho(1700)$ to the $B \to \eta_c{(1S,2S)} \pi\pi$
decays in the PQCD framework.
We calculated the branching ratios of the quasi-two-body decays $B \to \eta_c{(1S,2S)}
(\rho(770), \rho(1450),\rho(1700) \to ) \pi\pi$ by utilizing the vector current time-like form factor $F_\pi(s)$ with the
inclusion of the final state interactions between the pion pair in the resonant regions.

From the analytical analysis and the numerical results, we found the following points:
\begin{itemize}
\item[(1)]
The PQCD predictions for the branching ratios of the considered quasi-two-body decays are generally
in the order of $10^{-7}$ to $ 10^{-6}$.
We obtained the theoretical predictions for the branching ratios of the two-body decays
${\cal B}(B \to \eta_c{(1S,2S)} (\rho, \rho(1450),\rho(1700))$ out of the PQCD predictions
for the corresponding quasi-two-body decay modes, which will be tested
by future LHCb and Belle II experiments.

\item[(2)]
The whole pattern of the $\omega$-dependence of the pion form factor-squared  $|F_\pi|^2$ measured by
the {\it BABAR} Collaboration could be understood based on our studies, as illustrated in Fig.~\ref{fig2}(a).
The dominant contribution comes from the $\rho(770)$ resonance, while the deep dip around
$\omega \approx 1.6~{\rm GeV}$ is induced by the strong destructive interference between the
contribution from $\rho(1450)$ and $\rho(1700)$.

\item[(3)]
The general expectation based on the similarity between $B \to \eta_c \pi\pi$ and $B \to J/\psi \pi\pi$ decays
are confirmed: the value of newly defined ratio $R_2(\eta_c)\approx 0.45$ agrees well with the measured value
$R_2(J/\psi) =0.56\pm 0.09$ as reported by LHCb experiments.

\item[(4)]
The new ratios $R_3(\eta_c(1S))$ and $R_4(\eta_c(2S))$ among the branching ratios of the considered
decay modes are defined, and the PQCD predictions for their values will be tested by future experiments.
\end{itemize}

%-----------------------%
\begin{acknowledgments}

Many thanks to Hsiang-nan Li and Wen-Fei Wang for valuable discussions. This work is supported by
the National Natural Science Foundation of China under the Grant No.~11235005, ~11547020 and 11775117.
Ya Li and Ai-Jun Ma are also supported by the Project on Graduate Students Education and Innovation
of Jiangsu Province under Grant No. KYCX17-1057 and No. KYCX17-1056.

\end{acknowledgments}
%-----------------------%
\appendix

\section{Decay amplitudes}

The widely used wave function of  $B$ meson is adopted as the one being used in
Refs.~\cite{Keum:2000wi,prd65,epjc28-515,li2003,Xiao:2011tx,omega01,omega02},
\begin{eqnarray}
\Phi_B= \frac{i}{\sqrt{2N_c}} ({ p \hspace{-2.0truemm}/ }_B +m_B) \gamma_5 \phi_B ({\bf k_B}), \label{bmeson}
\end{eqnarray}
with the distribution amplitude
\begin{eqnarray}
\phi_B(x,b)&=& N_B x^2(1-x)^2\mathrm{exp} \left  [ -\frac{M_B^2\ x^2}{2 \omega_{B}^2} -\frac{1}{2} (\omega_{B}\; b)^2\right]
,  \label{phib}
\end{eqnarray}
where $N_B$ is the normalization factor defined through the normalization relation
$\int_0^1 dx \; \phi_B(x,b=0)=f_B/(2\sqrt{6})$. We also set $\omega_B = 0.40 \pm0.04$ GeV in the
numerical calculations.

For the final-state $\eta_c{(1S,2S)}$, its wave function can be written as
\begin{eqnarray}
\Psi_{\eta_c}&=&\frac{1}{\sqrt{2N_c}}{\gamma_5}\left[{ p \hspace{-2.0truemm}/ }_3\psi_v+m_{\eta_c}\psi_s\right],
\end{eqnarray}
where the twist-2 and twist-3 distribution amplitudes $\psi_{v}$ and $\psi_{s}$
for the $\eta_c(1S,2S)$ meson are parameterized as~\cite{epjc60-107,epjc75-293}
\begin{eqnarray}\label{eq:wave}
\psi_v(x,b)&=&\frac{f_{\eta_c}}{2\sqrt{6}}N^{v} x\bar{x}\; \mathcal {T}(x)
\cdot \exp\left [ -x\bar{x}\frac{m_c}{\omega}[\omega^2b^2+(\frac{x-\bar{x}}{2x\bar{x}})^2] \right ]\;,\non
\psi_s(x,b)&=&\frac{f_{\eta_c}}{2\sqrt{6}}N^s\; \mathcal {T}(x)
\cdot \exp\left [ -x\bar{x}\frac{m_c}{\omega}[\omega^2b^2+(\frac{x-\bar{x}}{2x\bar{x}})^2] \right]\;,
\end{eqnarray}
with the function ${\cal T }(x)=1$ for the meson $\eta_c(1S)$, and ${\cal T}(x)
=1-4b^2m_c\omega x\bar{x}+ m_c(x-\bar{x})^2/(\omega x\bar{x} )$ for the meson $\eta_c(2S)$.
The normalization constants $N^v$ and $N^s$ can be determined by the relation
$ \int_0^1\psi_{i}(x,b=0)d x = f_{\eta_c} /(2\sqrt{6})$.
The decay constant $f_{\eta_c(1S)}=0.42 \pm 0.05$ GeV and $\omega=0.6\pm 0.1$ GeV are
adopted for $\eta_c(1S)$ meson, while  $f_{\eta_c(2S)}=0.243^{+0.079}_{-0.111}$ GeV and
$\omega=0.2\pm 0.1$ GeV for $\eta_c(2S)$ meson.

The total decay amplitudes for the considered decay modes $B \to \eta_c{(1S,2S)} \pi\pi$ in
this work are given as follows:
\begin{eqnarray}
{\cal A}(B^+\to \eta_c{(1S,2S)}(\rho^+ \to )\pi^+\pi^0)&=&\frac{G_F}{\sqrt{2}}
\Big \{V^*_{cb}V_{cd} \left [F^{LL} + M^{LL}\right ]\non
&& \hspace{-3cm}
- V^*_{tb}V_{td} \Big [F^{\prime LL}+F^{LR} +M^{\prime LL} +M^{SP} \Big ]\Big\}\;,\\
{\cal A}(B^0\to \eta_c{(1S,2S)}(\rho^0 \to )\pi^+\pi^-)&=&-\frac{1}{\sqrt{2}}\mathcal{A}(B^+\to \eta_c{(1S,2S)}
(\rho^+ \to )\pi^+\pi^0)\;,
\end{eqnarray}
where $G_F=1.16639\times 10^{-5}$ GeV$^{-2}$ is the Fermi coupling constant and $V_{ij}$'s are the
Cabibbo-Kobayashi-Maskawa matrix elements.
The functions $ ( F^{LL}, F^{\prime LL}, F^{LR}, M^{LL}, M^{\prime LL}, M^{SP} ) $ appeared in above equations are
the individual decay amplitudes corresponding to different currents, the relevant Wilson coefficients have been included in
$F^{LL}$ and other functions, and their explicit expressions can also be found in Ref.~\cite{epjc76-675}.

%=============== Refs ===============%

\end{document}